\documentclass[preprint,pra,showpacs,nofootinbib]{revtex4-1}

\usepackage{amsmath,amssymb,amsthm}
\usepackage[mathscr]{euscript}
\usepackage{graphicx}
\usepackage{float}
\usepackage{dcolumn}
\usepackage{bm}
\usepackage{subfig}
\usepackage{hyperref}
\usepackage{xcolor}

\newcommand{\bea}{\begin{eqnarray}}
\newcommand{\eea}{\end{eqnarray}}
\newcommand{\beq}{\begin{equation}}
\newcommand{\eeq}{\end{equation}}

\begin{document}

\title{Optimal Quantum Metrology for Probing the Unruh Effect with Uniformly Accelerated Two-Level Atoms}
\author{  Yao Jin\footnote{Corresponding author. yaojin@gyu.edu.cn} }
\affiliation{School of Science,
Guiyang University,  Guiyang, Guizhou 550005, China}

\begin{abstract}

We develop a quantum metrological framework for optimizing the probing of the Unruh effect using uniformly accelerated two-level atomic probes. The acceleration-dependent factor generated during the atom--field interaction is encoded in the atomic state and can be estimated through repeated quantum measurements. For a fixed total probe time, which characterizes the available measurement resource, we optimize the interrogation time of individual probes, the initial atomic state, and the corresponding measurement basis to minimize the estimation uncertainty.
We show that the achievable precision is governed by the Fisher information accumulated per unit probe time. Under a fixed total probe time, shorter evolution times of individual probes allow for more sequential measurements, leading to a significant improvement in the estimation precision. The optimal initial state depends on both the acceleration factor and the probe evolution time. In particular, the excited state provides superior sensitivity in the weak-acceleration and small probe time regime. Furthermore, we determine the minimum total probe time required to resolve the acceleration-dependent signal associated with the Unruh effect and demonstrate that this requirement can be substantially reduced by employing atomic systems with larger transition dipole moments.
Our results establish an optimized quantum metrological strategy for probing acceleration-induced quantum effects and provide a systematic approach toward the experimental investigation of the Unruh effect.

\end{abstract}
\pacs{06.20.-f, 03.65.Yz, 03.65.Ta}

\maketitle

\section{Introduction}

Uniformly accelerated observers perceive the vacuum fluctuations of quantum fields defined in an inertial frame as a thermal bath with a temperature proportional to their acceleration~\cite{Fulling1973,Davies1975,Unruh1976}. This phenomenon, known as the Unruh effect, establishes a profound connection between quantum field theory in non-inertial frames and black hole physics, where a related mechanism leads to Hawking radiation~\cite{Hawking1974,Hawking1975}.
Experimental investigations of the Unruh effect typically rely on the response of a quantum detector coupled to fluctuating quantum fields. During the detector--field interaction, the detector state gradually acquires acceleration-dependent modifications. However, the corresponding acceleration-induced signatures, such as thermal excitation contributions, are extremely small under realistic experimental conditions, making direct observation of the Unruh effect a major challenge.

Quantum metrology provides a powerful framework for extracting weak physical signals from quantum systems through optimized state preparation and measurement strategies. The outcomes of repeated measurements on quantum probes follow a probability distribution determined by the parameter of interest. By constructing an optimal estimator, the parameter can be inferred with an uncertainty bounded by the Cram\'{e}r--Rao inequality~\cite{Helstrom,Holevo,Hubner,Braunstein}. The attainable precision is governed by the Fisher information, which quantifies the information extracted from quantum measurements and determines the ultimate sensitivity of parameter estimation.
Quantum metrology has been successfully applied to various precision measurement tasks, including quantum frequency standards~\cite{Bollinger}, optimal quantum clocks~\cite{Buzek}, measurements of gravitational acceleration~\cite{Peters}, and clock synchronization~\cite{Jozsa}. These developments motivate a quantum sensing approach to investigating relativistic quantum effects. In particular, the acceleration-dependent modification encoded in the state of a uniformly accelerated atomic probe can be extracted through repeated quantum measurements. To resolve the Unruh-induced signal, the uncertainty of the estimated acceleration-dependent parameter must be smaller than its magnitude. Therefore, under a fixed total probe time, the available experimental resource, it is essential to optimize the information extracted per unit time and determine the minimum measurement resource required for probing the Unruh effect.

In this work, we formulate the observation of the Unruh effect as a quantum metrological problem and investigate an optimized sensing strategy using uniformly accelerated two-level atoms coupled to the vacuum electromagnetic field. We first analyze the atomic evolution and identify the acceleration-dependent contribution induced by the atom--field interaction. By optimizing the interrogation time, initial atomic state, and measurement basis of each probe, we derive the ultimate estimation precision under a fixed total probe time and determine the optimal probing strategy. Our results demonstrate how quantum metrological optimization can substantially reduce the experimental resource requirements for investigating acceleration-induced quantum effects.

\section{Evolution of the probe two-level atom coupled to fluctuating electromagnetic field in the multipolar coupling scheme}
We work in the weak-coupling and Markovian regime, where the atom-field coupling is sufficiently small and the relevant field correlation functions decay on timescales much shorter than the atomic evolution timescale. The validity of the Lindblad master equation for accelerating detectors has been discussed in the literature; it is reliable in the weak-coupling limit, while non-Markovian corrections may become important outside this regime~\cite{JHEP2020,PRD2025}. In this work, we therefore restrict our analysis to the regime \(\gamma_0/\omega_0 \ll 1\) and to probe times for which memory effects are negligible.
We begin by review the evolution of a two-level atom coupled to fluctuating electromagnetic field in the multipolar coupling scheme.
The total Hamiltonian of a two-level polarizable atom and the electromagnetic field has the form
$
H=H_s+H_f+H_I
$.
Here $H_s={1\over
2}\,\hbar\omega_0\sigma_3$ denotes the Hamiltonian of
the two-level atom with $\omega_0$ denoting the transition frequency. $H_f$ denotes the
Hamiltonian of the free electromagnetic field. $H_I$ denotes the the interaction Hamiltonian
between the polarized two-level atom and the electromagnetic field. In the multipolar
coupling scheme, $H_I$ has the form
\begin{equation}\label{HI}
 H_I(\tau)=-e\textbf{r} \cdot
\textbf{E}(x(\tau))\;,
\end{equation}
with {\it e} denoting the electron
electric charge, $e\,\bf r$  denoting the atomic electric dipole moment, and
${\bf E}(x)$ denoting the electric field strength.

We assume the electromagnetic field is initially in vacuum state from the aspect of inertial observer. So the initial total density matrix $\rho_{tot}(0)=\rho(0) \otimes |0\rangle\langle0|$ with $\rho(0)$ denoting the initial reduced density matrix of the atom, and $|0\rangle$ denoting the vacuum state of the field. The evolution of the total density matrix $\rho_{tot}$ at time $\tau$ satisfies
\begin{equation}\label{evo}
\frac{\partial\rho_{tot}(\tau)}{\partial\tau}=-\frac{i}{\hbar}[H,\rho_{tot}(\tau)]\;.
\end{equation}
In the weak coupling assumption, the evolution of the reduced
density matrix $\rho(\tau)$ is given in the
Kossakowski-Lindblad form~\cite{Lindblad, pr5}
\begin{equation}\label{master}
{\partial\rho(\tau)\over \partial \tau}= -\frac{i}{\hbar}\big[H_{\rm eff},\,
\rho(\tau)\big]
 + {\cal L}[\rho(\tau)]\ ,
\end{equation}
where
\begin{equation}
{\cal L}[\rho]={1\over2} \sum_{i,j=1}^3
a_{ij}\big[2\,\sigma_j\rho\,\sigma_i-\sigma_i\sigma_j\, \rho
-\rho\,\sigma_i\sigma_j\big]\ .
\end{equation}
The coefficients of the Kossakowski matrix $a_{ij}$ are written as
\begin{equation}
a_{ij}=A\delta_{ij}-iB
\epsilon_{ijk}\delta_{k3}-A\delta_{i3}\delta_{j3}\;,
\end{equation}
with
\begin{equation}\label{abc}
A=\frac{1}{4}[{\cal {G}}(\omega_0)+{\cal{G}}(-\omega_0)]\;,\;~~
B=\frac{1}{4}[{\cal {G}}(\omega_0)-{\cal{G}}(-\omega_0)]\;.
\end{equation}
Here
\begin{equation}
{\cal G}(\lambda)=\int_{-\infty}^{\infty} d\Delta\tau \,
e^{i{\lambda}\Delta\tau}\, G^{+}\big(\Delta\tau\big)
\;,
\end{equation}
with $G^{+}(x-x')$ being shown in relation with the two-point functions of the electromagnetic fields $\langle0|E_i(x)E_j(x')|0 \rangle$ as
\begin{equation}
G^{+}(x-x')=\frac{e^2}{\hbar^2} \sum_{i,j=1}^3\langle +|r_i|-\rangle\langle -|r_j|+\rangle\,\langle0|E_i(x)E_j(x')|0 \rangle\;.
\end{equation}
By Absorbing the Lamb shift term, the effective Hamiltonian $H_{\rm eff}$ is written as
\begin{equation}\label{heff}
H_{\rm eff}=\frac{1}{2}\hbar\Omega\sigma_3={1\over 2}\hbar\{\omega_0+{i\/2}[{\cal
K}(-\omega_0)-{\cal K}(\omega_0)]\}\,\sigma_3\;,
\end{equation}
with $\Omega$ denoting the effective level spacing of the atom, and
\begin{equation}
{\cal K}(\lambda)=\frac{P}{\pi
i}\int_{-\infty}^{\infty} d\omega\ \frac{ {\cal G}(\omega)
}{\omega-\lambda}\;.
\end{equation}
%Let us expand the density matrix $\rho$ in terms of the Pauli matrices,
%\begin{equation}\label{density}
%\rho({\tau})=\frac{1}{2}\bigg(1+\sum_{i=1}^{3}\rho_i({\tau})\sigma_i\bigg)\;.
%\end{equation}
%Plugging Eq.~(\ref{density}) into Eq.~(\ref{master})
Let us expand the density matrix $\rho(\tau)$ in terms of the Pauli matrices,
\begin{equation}\label{density}
\rho(\tau)=\frac{1}{2}\bigg(I+\sum_{i=1}^{3}n_i(\tau)\sigma_i\bigg) = \frac{1}{2}[I+\mathbf{n}(\tau)\cdot\boldsymbol{\sigma}]\;,
\end{equation}
where \(\mathbf{n}(\tau)=(n_1(\tau),n_2(\tau),n_3(\tau))\) denotes the Bloch vector.

The initial state of the two-level atom in each probe is assumed to be prepared in pure state as \(\cos\frac{\theta}{2}|+\rangle+\sin\frac{\theta}{2}e^{i\phi}|-\rangle\) with \(\theta\), and \(\phi\) denoting the weight and phase factor.
Therefore, the Bloch vector with proper time \(\tau\) is calculated as:
\begin{eqnarray}\label{Bloch}
&&n_1(\tau)=\sin\theta\cos(\Omega\tau+\phi)\,e^{-2A\tau}\;,\nonumber\\
&&n_2(\tau)=\sin\theta\sin(\Omega\tau+\phi)\,e^{-2A\tau}\;,\\
&&n_3(\tau)=\cos\theta\, e^{-4A\tau}-\frac{B}{A}(1-e^{-4A\tau})\nonumber\;.
\end{eqnarray}
As a result, the evolution of the state of the two-level atom is determined by the factors $A$ and $B$, which are determined by the field correlation function. Therefore, the distribution of field modes as well as the initial state of the field in the frame of the atom affect the evolution of the atom.

\section{Optimal estimation of the acceleration factor under a fixed total probe time}
The probe two-level atom is assumed to be move along the $x$-direction, with acceleration $a$.
Applying the trajectory of the atom
\begin{eqnarray}\label{traj}
t(\tau)=\frac{c}{a}\mathrm{sinh} \frac{a\tau}{c}\;, \ \ \ x(\tau)=\frac{c^2}{a}\mathrm{cosh} \frac{a\tau}{c}\;, \ \ \
y(\tau)=0\;,\ \ \ z(\tau)=0\;,
\end{eqnarray}
on the electric two-point function~\cite{Greiner}
\begin{eqnarray}\label{2p2}
\langle0|E_i(x(\tau))E_j(x(\tau'))|0\rangle &=& \frac{\hbar c}{4\pi^2 \varepsilon_0}(\partial_0\partial_0^\prime\delta_{ij}-\partial_i\partial_j^\prime)\nonumber\\
&&\times \frac{1}{(x-x')^2+(y-y')^2+(z-z')^2-(ct-ct'-i\varepsilon)^2}\;,\label{ee1}
\end{eqnarray}
where
$\varepsilon\rightarrow+0$, $\partial^\prime$ denotes the
differentiation with respect to $x^\prime$.
The electric-field two-point functions in the frame of atom can be calculated as~\cite{em1,em2,em3}
\begin{eqnarray}
\langle0|E_i(x(\tau))E_j(x(\tau'))|0\rangle
=\frac{\hbar}{16\pi^2\varepsilon_0c^7}\frac{a^4\delta_{ij}}{\mathrm{sinh}^4[\frac{a}{2c}(\tau-\tau'-i\varepsilon)]}\;,
\end{eqnarray}
and its corresponding Fourier transforms become
\begin{eqnarray}\label{fourier0}
{\cal G}(\lambda)=\sum_i
\frac{e^2|\langle -|r_i|+\rangle|^2\lambda^3}{6\pi\varepsilon_0\hbar c^3}
\bigg(1+\frac{a^2}{c^2\lambda^2}\bigg)
\bigg(1+\mathrm{coth}\frac{\pi c\lambda}{a}\bigg)
\;,
\end{eqnarray}
with $\theta(\lambda)$ being the standard step function. Here we let $\varepsilon=0$ after the calculation.
All coefficients that affect the evolution of the atom are calculated as
\begin{equation}
A=\frac{1}{4}\gamma_0\bigg(1+X^2\bigg)\frac{e^{2\pi/X}+1}{e^{2\pi/X}-1}\;,
\end{equation}
\begin{equation}
B=\frac{1}{4}\gamma_0\bigg(1+X^2\bigg)\;,
\end{equation}
\begin{equation}\label{lm}
\Omega=\omega_0+\frac{\gamma_0}{2\pi\omega_0^3}\,P\int_0^\infty
d\omega\,\omega^3\left(\frac{1}{\omega+\omega_0}-\frac{1}{\omega-\omega_0}\right)\left(1+X^2\right)\left(1+\frac{2}{e^{2\pi/X}-1}\right)\,.
\end{equation}
Here $\gamma_0=e^2|\langle -|{\bf
r}|+\rangle|^2\,\omega_0^3/3\pi\varepsilon_0\hbar c^3$ denotes the spontaneous emission rate in Minkowski-vacuum, and $X=\frac{a}{c\omega_0}$ denotes the dimensionless acceleration parameter to be estimated.
Apparently, the evolution of the atom becomes acceleration-dependent. To estimate the acceleration effect, measurement is performed on the state of the atom.
Without loss of generality, the measurement is assumed in the basis of
\(\{\cos[\frac{\alpha}{2}+i\frac{\pi}{2}]|+\rangle+e^{i\beta}\sin[\frac{\alpha}{2}+i\frac{\pi}{2}]|-\rangle\}\), \((i=0,1)\).
Here \(\alpha\) and \(\beta\) are arbitrary weight and phase factor of the basis of measurement, and \(i\) denotes the result of measurement.
Denoting these basis states as \(|M_i\rangle\), the probability for outcome \(i\) is \(P_i = \mathrm{Tr}[\rho(\tau)|M_i\rangle\langle M_i|]\). Using the Bloch representation \(\rho(\tau)=\frac{1}{2}[I+\mathbf{n}(\tau)\cdot\boldsymbol{\sigma}]\), expanding this trace in the Pauli basis directly yields
\begin{equation}
P_i=\frac{1}{2}[1+(-1)^in_1(\tau)\sin\alpha\cos\beta+(-1)^in_2(\tau)\sin\alpha\sin\beta+(-1)^in_3(\tau)\cos\alpha]\;.
\end{equation}
The above procedure including the preparation of initial atomic state, atom-field evolution and measurement can be repeated $N$ times.
Then using maximum-likelihood estimation, the acceleration parameter $X$ can be estimated with the uncertainty of the estimation satisfying~\cite{Helstrom,Holevo,Hubner,Braunstein}
\begin{equation}
Var (X)\geq\frac{1}{N F_X}
\end{equation}
with
\begin{equation}\label{F}
F_X=\sum_i\frac{(\partial_X P_i)^2}{P_i}=\frac{[\partial_X n_1(\tau)\sin\alpha\cos\beta+\partial_X n_2(\tau)\sin\alpha\sin\beta+\partial_X n_3(\tau)\cos\alpha]^2}{1-[n_1(\tau)\sin\alpha\cos\beta+n_2(\tau)\sin\alpha\sin\beta+n_3(\tau)\cos\alpha]^2}\;.
\end{equation}
We let $T$ denotes the total time of $N$ probes, $\tau$ denotes the time of each probe. Therefore, $N=\frac{T}{\tau}$, and the uncertainty of the estimation satisfies
\begin{equation}
Var (X)\geq\frac{1}{N F_X}=\frac{1}{T F_X/\tau}=\frac{1}{T\gamma_0 F_X/\widetilde{\tau}}\;
\end{equation}
with $\widetilde{\tau}=\gamma_0\tau$ denotes the dimensionless proper time in one probe.
To test Unruh effect, $\Delta X=\sqrt{Var (X)}$ should be smaller than $X$, which indicates the total probe time $T$ should be large enough. Therefore, the minimal $T$ required to test Unruh effect is determined by the largest $F_X/\widetilde{\tau}$.
Therefore, the evolution time of each probe, the initial state in each proper, and the basis of measurement should be optimized to achieve the largest $F_X/\widetilde{\tau}$.

We would like to clarify that the measurement parameters \(\alpha\) and \(\beta\) are continuous variables, and for each given initial state, we should optimize them to maximize the classical Fisher information. The maximum classical Fisher information is usually dubbed as the quantum Fisher information (QFI).

The QFI and the optimal basis of measurement can be deduced through the symmetric logarithmic derivative (SLD) \(L_X\) for a parameter \(X\), which is defined implicitly by
$
\partial_X \rho = \frac{1}{2}(L_X \rho + \rho L_X)
$.
Since \(L_X\) is a Hermitian operator acting on a two-level system, it can be expanded in the Pauli basis as
$
L_X = l_0 I + \mathbf{l} \cdot \boldsymbol{\sigma}
$,
where \(l_0\) is a real scalar and \(\mathbf{l}\) is a real vector.

Substituting these into the SLD definition, the vector \(\mathbf{l}\) is solved as
\begin{equation}
\mathbf{l} = \dot{\mathbf{n}} + \frac{\mathbf{n} \cdot \dot{\mathbf{n}}}{1 - |\mathbf{n}|^2} \mathbf{n},
\end{equation}
with \(\dot{\mathbf{n}} = \partial_X \mathbf{n}\). The scalar part is \(l_0 = -\mathbf{l} \cdot \mathbf{n}\). When the Bloch vector of the measurement basis \((-1)^i (\sin\alpha\cos\beta, \sin\alpha\sin\beta, \cos\alpha)\) becomes parallel to \(\mathbf{l}\), the classical Fisher information achieves to its maximum value. Then the quantum Fisher information is given by
$
F_X = \mathrm{Tr}(\rho L_X^2)
$.
Using the explicit form of \(L_X\), one can verify that~\cite{Zhong}
\begin{equation}
F_X = |\mathbf{l}|^2 - (\mathbf{n} \cdot \mathbf{l})^2 = |\dot{\mathbf{n}}|^2 + \frac{(\mathbf{n} \cdot \dot{\mathbf{n}})^2}{1 - |\mathbf{n}|^2}.
\end{equation}
In the pure-state limit \(|\mathbf{n}| \to 1\), we have \(\mathbf{n} \cdot \dot{\mathbf{n}} = 0\), so that \(\mathbf{l} = \dot{\mathbf{n}}\) and \(F_X = |\dot{\mathbf{n}}|^2\). In the following discussion, we use the optimal measurement basis, and therefore, the classical Fisher information becomes the quantum Fisher information.

\begin{figure}[htbp]
\centering
\subfloat[Small \(\widetilde{\tau}\)]{\includegraphics[height=2.1in,width=3.1in]{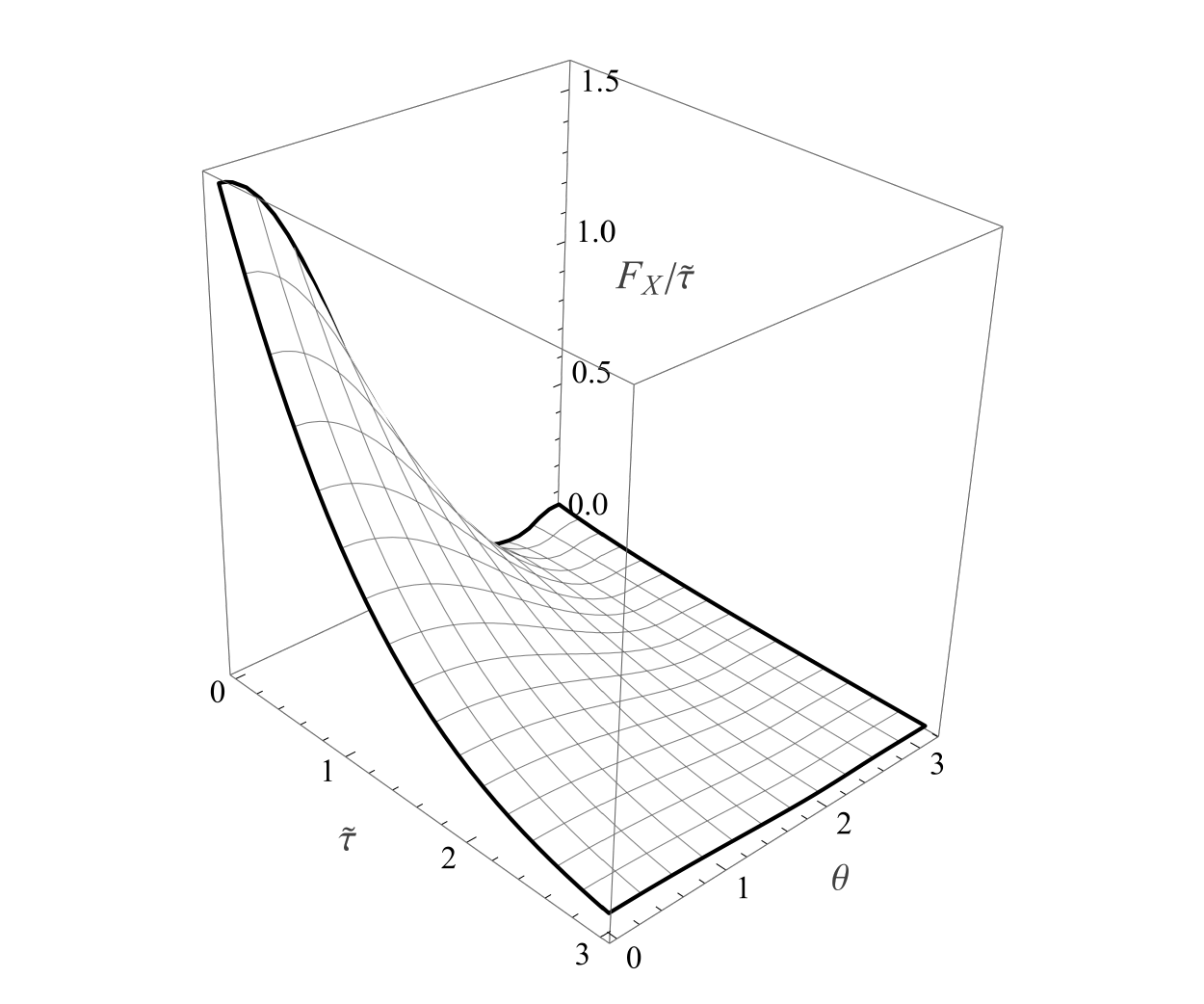}}
\hfill
\subfloat[Large \(\widetilde{\tau}\)]{\includegraphics[height=2.1in,width=3.1in]{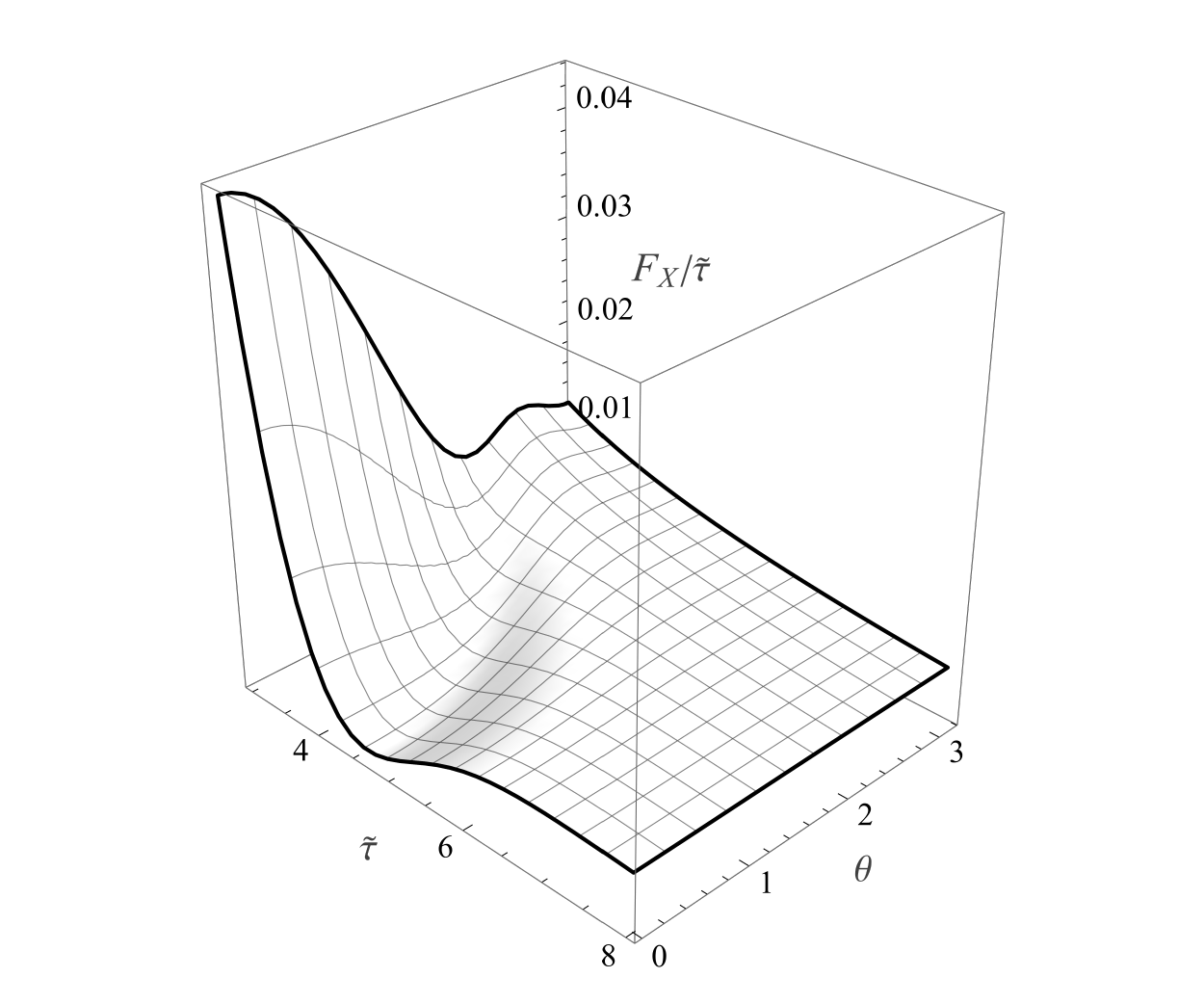}}
\caption{$F_X/\widetilde{\tau}$ as  a function of $\widetilde{\tau}$ and $\theta$ when \(X=0.8\).
}
\label{F1}
\end{figure}

\begin{figure}[htbp]
\centering
\subfloat[Small \(\widetilde{\tau}\)]{\includegraphics[height=2.1in,width=3.1in]{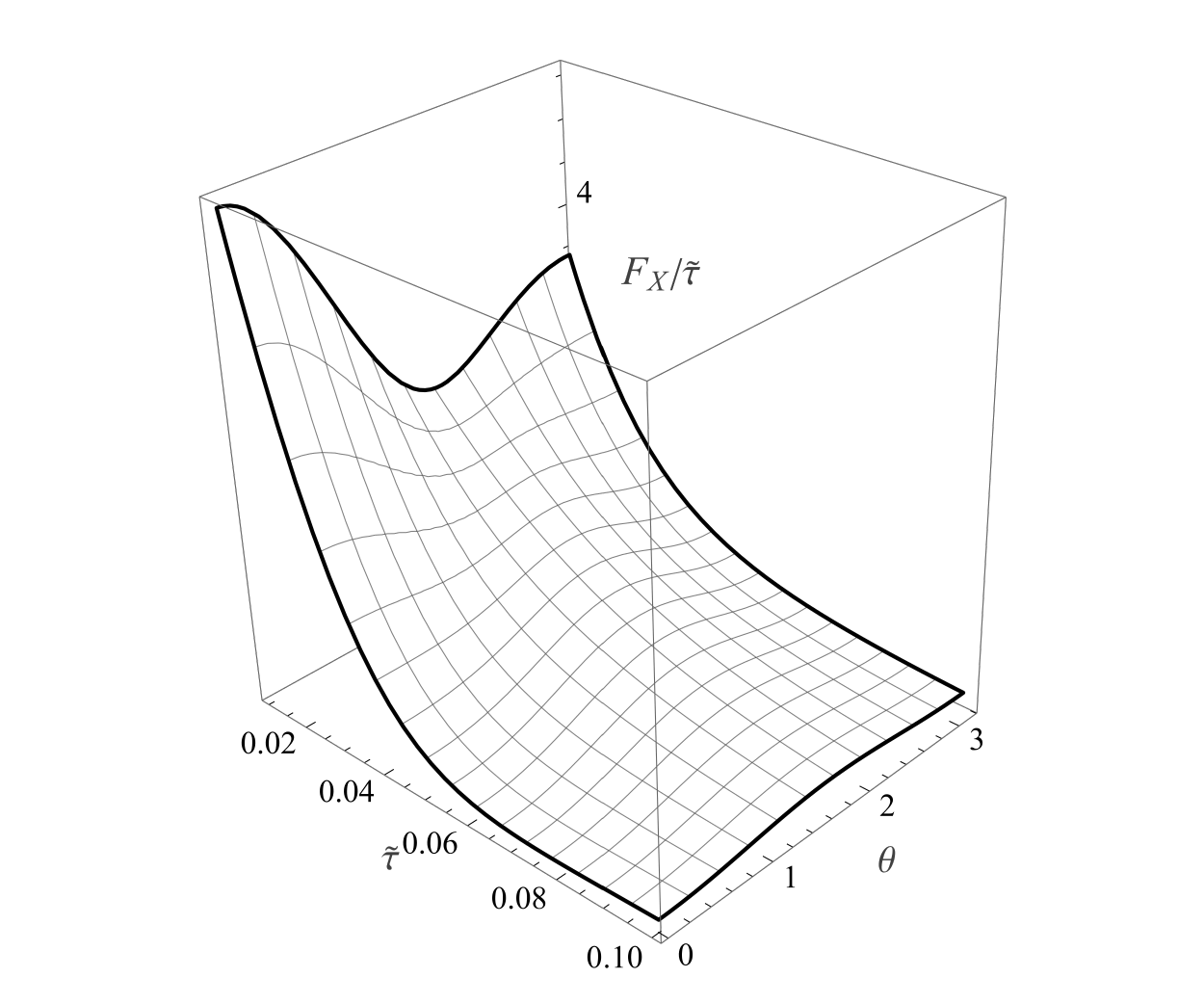}}
\hfill
\subfloat[Large \(\widetilde{\tau}\)]{\includegraphics[height=2.1in,width=3.1in]{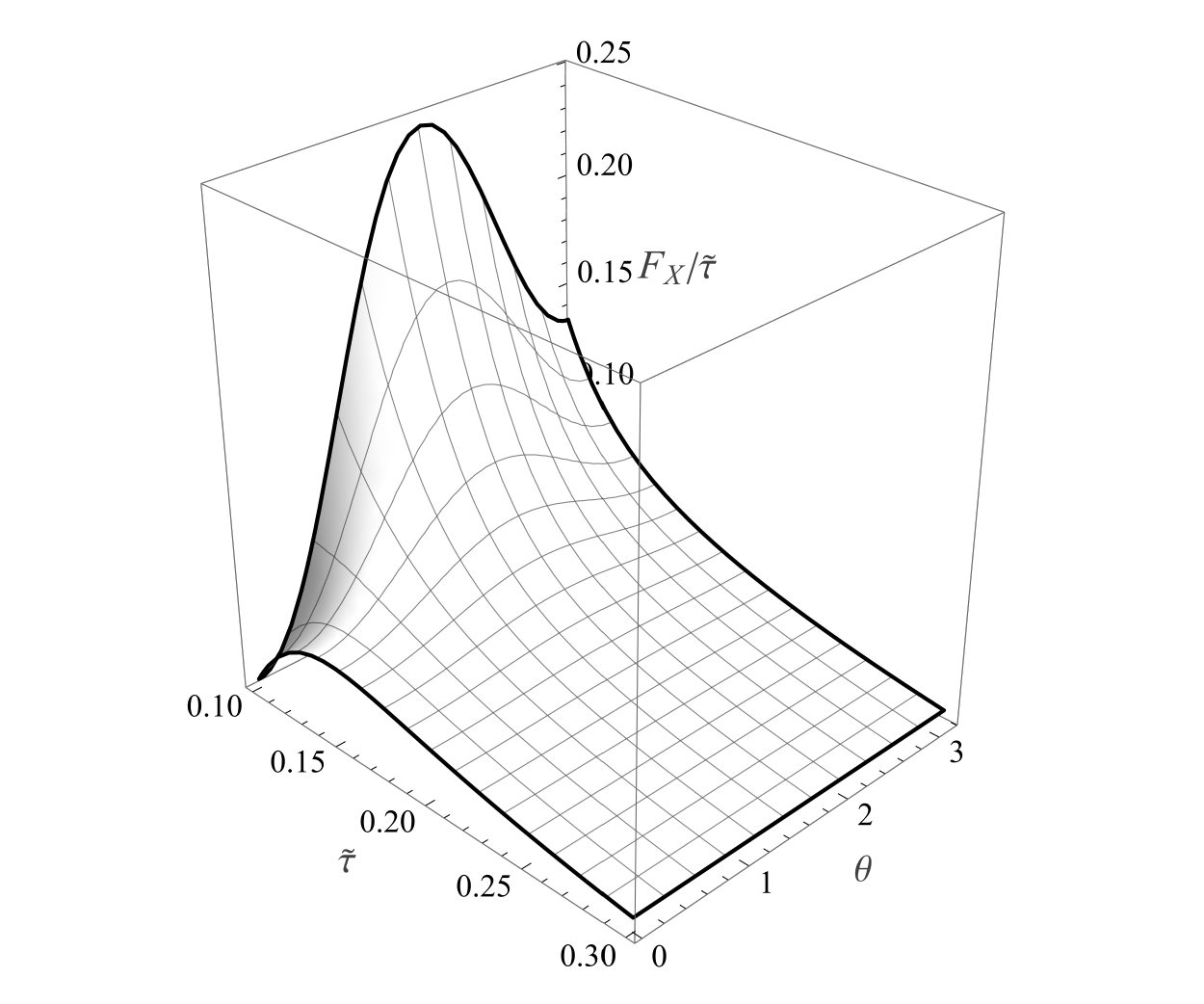}}
\caption{$F_X/\widetilde{\tau}$ as  a function of $\widetilde{\tau}$ and $\theta$ when \(X=5\).
}
\label{F2}
\end{figure}

\begin{figure}[htbp]
\centering
\subfloat[Small \(\widetilde{\tau}\)]{\includegraphics[height=2.1in,width=3.1in]{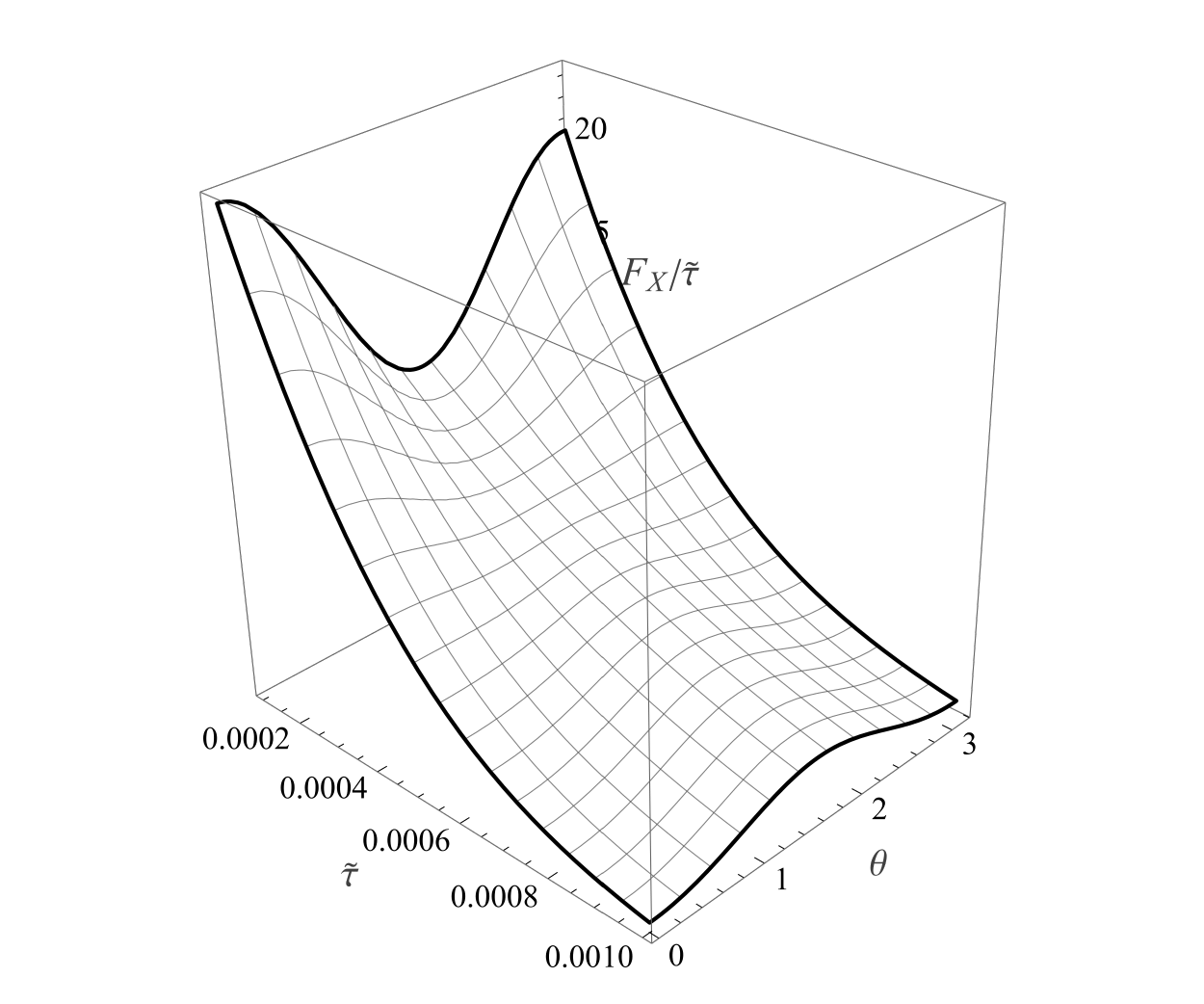}}
\hfill
\subfloat[Large \(\widetilde{\tau}\)]{\includegraphics[height=2.1in,width=3.1in]{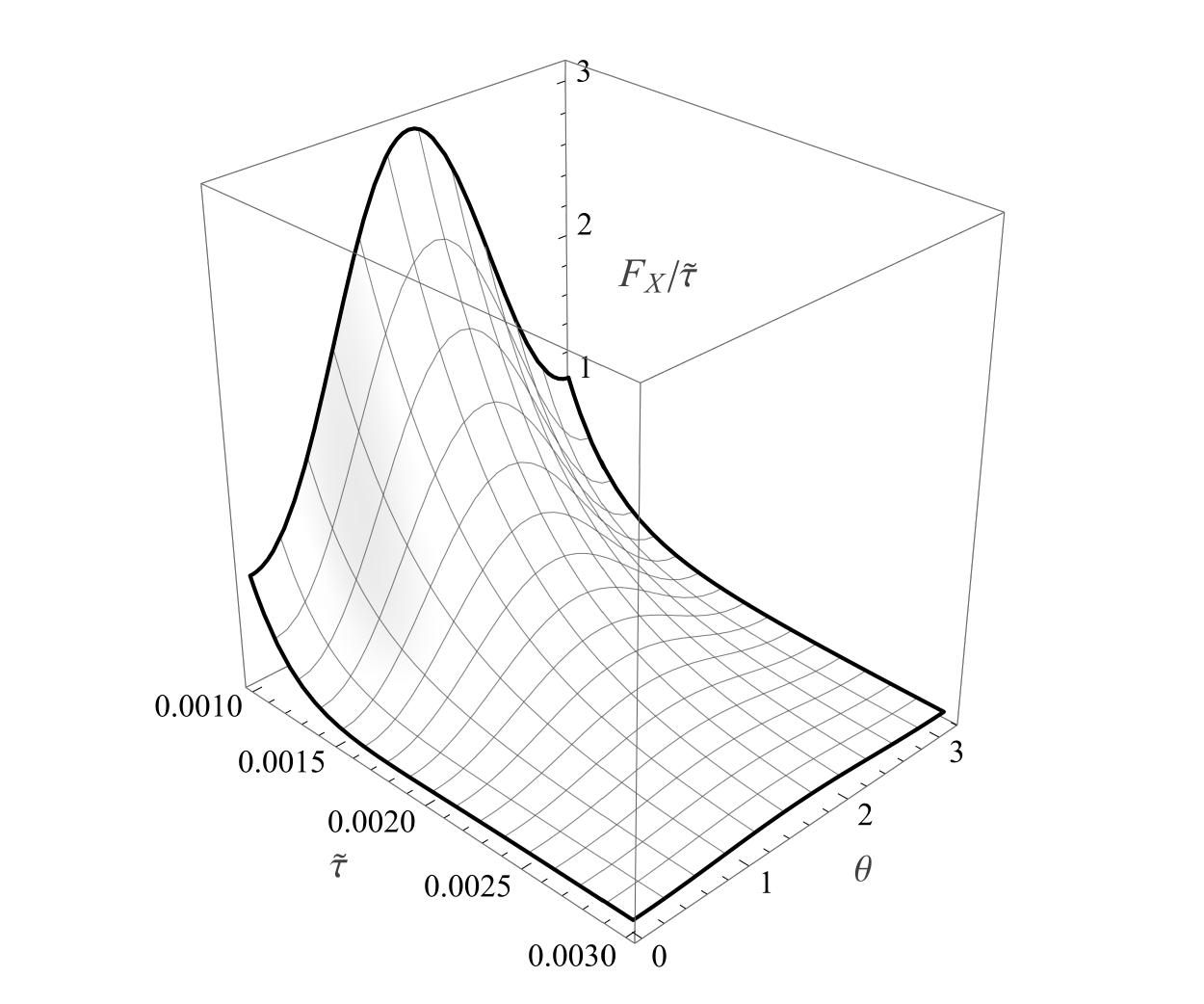}}
\caption{$F_X/\widetilde{\tau}$ as  a function of $\widetilde{\tau}$ and $\theta$ when \(X=20\).
}
\label{F3}
\end{figure}

\begin{figure}[htbp]
\centering
\subfloat[$\widetilde{\tau}=0.01$]{%
    \includegraphics[height=2.1in,width=3.1in]{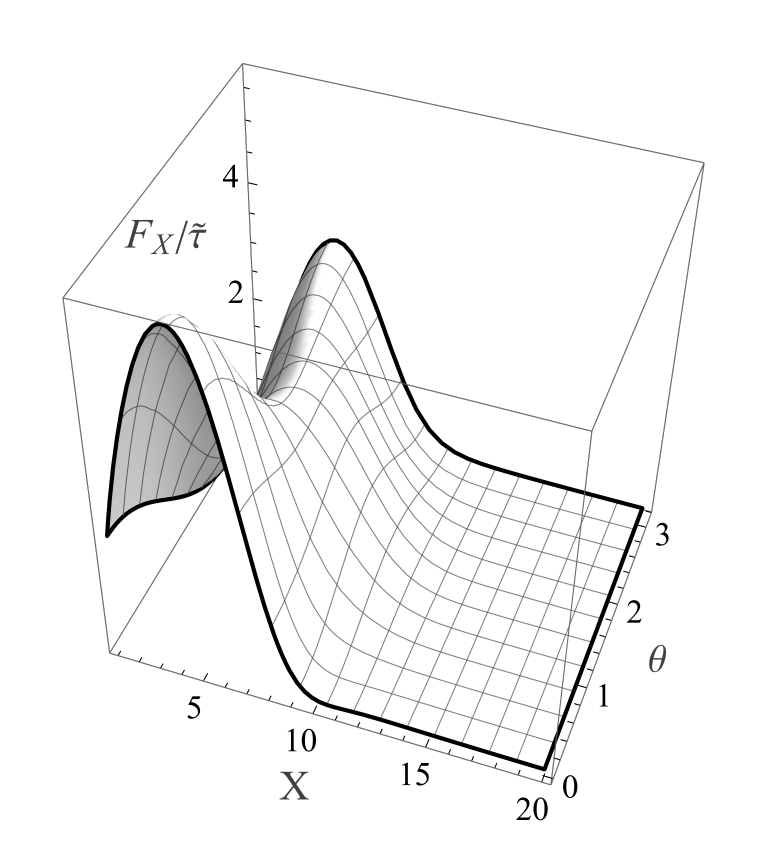}%
}%
\hfill
\subfloat[$\widetilde{\tau}=5$]{%
    \includegraphics[height=2.1in,width=3.1in]{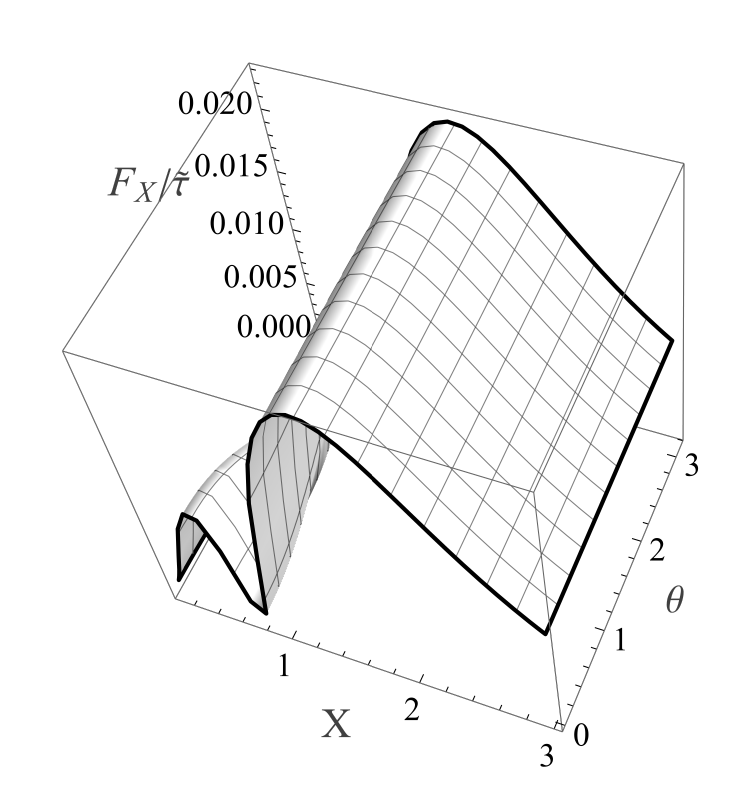}%
}%
\caption{$F_X/\widetilde{\tau}$ as a function of $X$ and $\theta$ at $\widetilde{\tau}=0.01$ and $5$.}
\label{F4}
\end{figure}

To optimize the evolution time and the initial state in each probe, we plot $F_X/\widetilde{\tau}$ in function of $\widetilde{\tau}$ and $\theta$ in small acceleration case with $X=0.8$, medium acceleration case with $X=5$, and large acceleration case with $X=20$ in Fig.~(\ref{F1}), Fig.~(\ref{F2}), Fig.~(\ref{F3}) respectively. The initial state factor $\theta$ ranges from $0$ to $\pi$. $\theta=0$ corresponds to the excited state $|+\rangle$, $\theta=\pi$ corresponds to the ground state $|-\rangle$, and $\theta=\frac{\pi}{2}$ corresponds to the equal weight superposition states.
Results show that, the optimal initial state is determined by the evolution time in each probe $\widetilde{\tau}$ as well as the magnitude of acceleration $X$.
For arbitrary acceleration, the excited state shows advantage in the small $\widetilde{\tau}$ cases, and the superposition states show advantage in the large $\widetilde{\tau}$ cases. The excited state becomes the optimal initial state when $\widetilde{\tau}$ is sufficiently small. The maximum $F_X/\widetilde{\tau}$ is obtained by choosing the excited state as the initial state and making the evolution time in each probe $\widetilde{\tau}$ approaches zero. The maximum $F_X/\widetilde{\tau}$ when $\widetilde{\tau}\rightarrow0$ becomes larger in cases with larger acceleration $X$. However, $F_X/\widetilde{\tau}$ decreases faster with the increase of $\widetilde{\tau}$. For a fixed probe time $\widetilde{\tau}$, to analyze the relation between $F_X/\widetilde{\tau}$ and acceleration factor $X$ for arbitrary initial state, we plot $F_X/\widetilde{\tau}$ as function of $X$ and $\theta$ in Fig.~(\ref{F4}) with $\widetilde{\tau}=0.01$ and $\widetilde{\tau}=5$ respectively.
Results show that for a fixed probe time $\widetilde{\tau}$, there always exists an optimal $X$ to reach the largest $F_X/\widetilde{\tau}$. In case of small probe time $\widetilde{\tau}=0.01$, the case with initial excited state outperforms other cases in the small acceleration condition $X<7.3$, while the case with initial superposition states take advantage in the large acceleration condition $X>7.3$. The maximum value of $F_X/\widetilde{\tau}$ with initial excited state becomes the largest among the cases with different initial states, and is reached at $X=4.4$, which is smaller than those with other initial states. Since small acceleration $X$ is easier to obtain in reality, the excited state becomes the optimal initial state in the small probe time case. However, in large probe time case as $\widetilde{\tau}=5$, oscillatory behavior occurs in case with initial excited state, and the largest value of $F_X/\widetilde{\tau}$ with arbitrary initial states become almost the same.

\begin{figure}[htbp]
\centering
\subfloat[\(X = 0.8\), \(\tilde{\tau} = 1\)]{\includegraphics[width=0.45\textwidth]{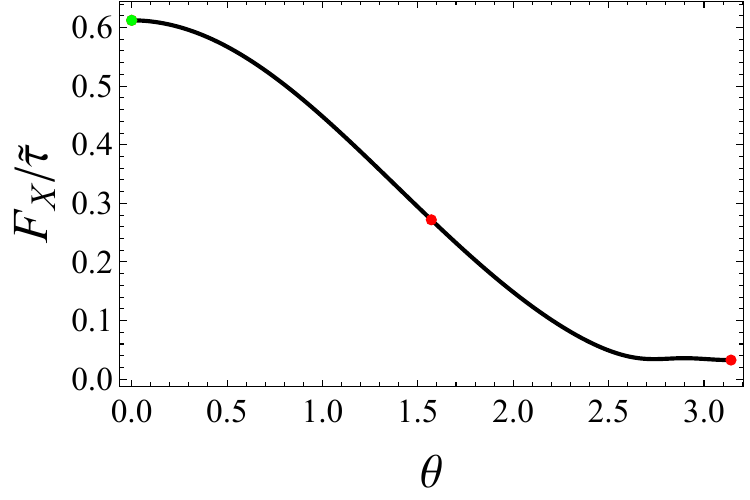}}
\hfill
\subfloat[\(X = 5\), \(\tilde{\tau} = 0.05\)]{\includegraphics[width=0.45\textwidth]{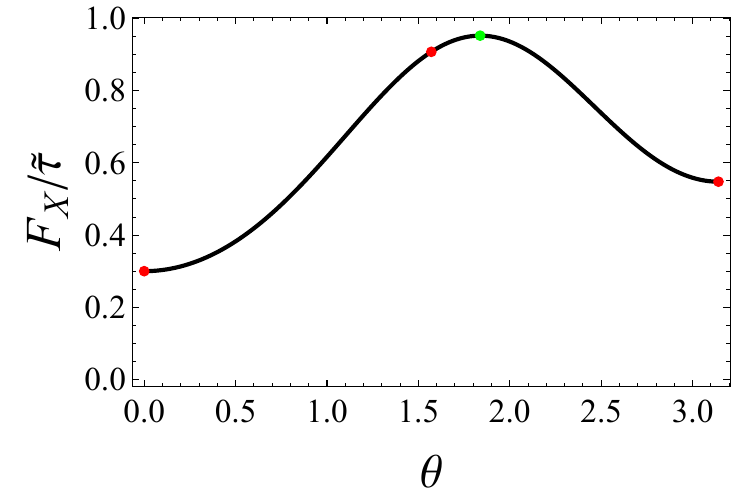}}
\\
\subfloat[\(X = 5\), \(\tilde{\tau} = 0.15\)]{\includegraphics[width=0.45\textwidth]{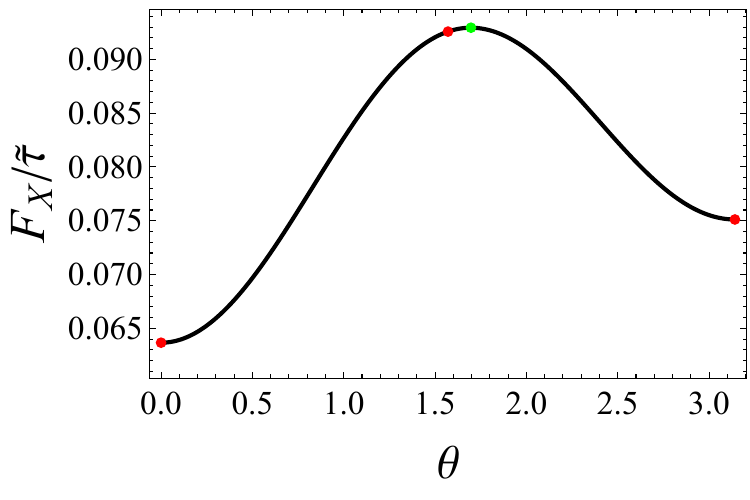}}
\hfill
\subfloat[\(X = 20\), \(\tilde{\tau} = 0.01\)]{\includegraphics[width=0.45\textwidth]{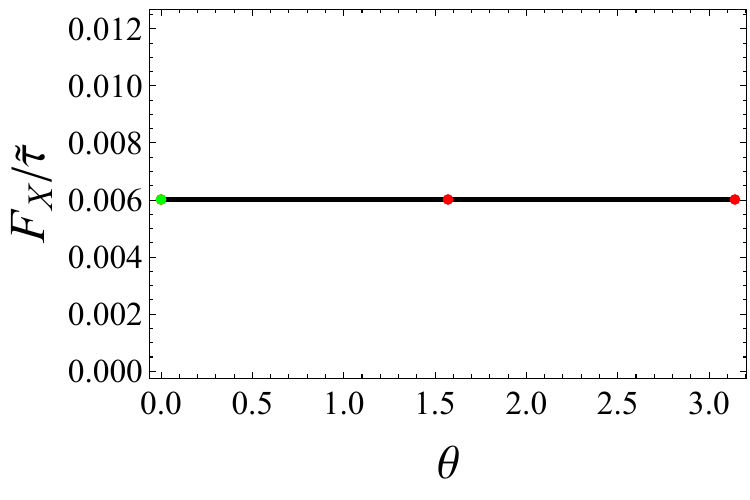}}
\caption{\(F_X/\tilde{\tau}\) as a function of the initial state parameter \(\theta\) for four different parameter regimes. The green dots indicate the strict global maxima, while the red dots mark the three representative initial states: \(\theta=0\) (excited), \(\theta=\pi/2\) (equal weight superposition), and \(\theta=\pi\) (ground).}
\label{Fig5}
\end{figure}

To show the optimal initial states exactly in explicit cases, for given $\widetilde{\tau}$ and $X$, we plot $F_X/\widetilde{\tau}$ as function of initial state parameter \(\theta\). The results are summarized in Fig.~\ref{Fig5}. As clearly shown in Fig.~\ref{Fig5}, the optimal initial state is highly parameter-dependent: In the weak acceleration regime shown in Fig.~\ref{Fig5}(a), the excited state is strictly optimal. For intermediate parameters with short probe times shown in Fig.~\ref{Fig5}(b), a general initial state (\(\theta \approx 1.8\) rad) provides a higher Fisher information than any representative state. For longer probe times shown in Fig.~\ref{Fig5}(c), the optimal initial state moves toward the equal weight superposition state. For strong acceleration and enough probe times shown in Fig.~\ref{Fig5}(d), the probe states are nearly thermalized, and $F_X/\widetilde{\tau}$ becomes independent of the initial state chosen.

\subsection*{Experimental feasibility and resource estimate}
To resolve the acceleration-dependent signal, the estimation uncertainty must satisfy \(\Delta X < X\). According to the Cram\'{e}r-Rao bound, \(\Delta X \ge 1/\sqrt{N F_X}\). The total probe time is \(T = N \tau = N \tilde{\tau}/\gamma_0\), where \(N\) is the number of probes and \(\tilde{\tau} = \gamma_0\tau\). Therefore, the minimum total measurement resource required to detect the signal is given by
\begin{equation}
T\gamma_0 > \frac{1}{X^2\,(F_X/\tilde{\tau})}.
\end{equation}
The dimensionless acceleration parameter is \(X = a/(c\omega_0)\). We consider two realistic atomic systems to estimate the required total probe time.

First, for the hydrogen atom \(2p\to1s\) transition, we use \(\omega_0 \sim 10^{15}\,\mathrm{Hz}\) and \(\gamma_0 \sim 10^{8}\,\mathrm{s^{-1}}\). For a physically accessible acceleration \(a \sim 10^{18}\,\mathrm{m/s^2}\), the acceleration factor is \(X \sim 10^{-5}\). Using the optimal Fisher information per unit probe time \(F_X/\tilde{\tau} \sim 1.3567\times 10^{-11}\), we obtain \(T\gamma_0 \sim 7.37\times 10^{20}\) and a required total probe time of \(T \sim 7.37\times 10^{12}\,\mathrm{s}\).

Second, for the cesium atom \(6P_{3/2}\to6S_{1/2}\) transition, we use \(\omega_0 \sim 10^{14}\,\mathrm{Hz}\) and \(\gamma_0 \sim 10^{8}\,\mathrm{s^{-1}}\). The lower transition frequency results in a larger acceleration factor \(X \sim 10^{-4}\). This increases the Fisher information to \(F_X/\tilde{\tau} \sim 1.3567\times 10^{-9}\), significantly reducing the required total probe time to \(T\gamma_0 \sim 7.37\times 10^{16}\) and \(T \sim 7.37\times 10^{8}\,\mathrm{s}\).

These estimates are order-of-magnitude calculations and assume idealized state preparation and measurement. Experimental imperfections such as finite measurement efficiency are not considered. Since the required time is highly sensitive to the atomic transition parameters, employing atomic systems with larger transition dipole moments and tunable transition frequencies, such as Rydberg atoms, can further substantially reduce the required total probe time.

\section{Conclusion}

In summary, we have developed a quantum metrological framework for optimizing the probing of the Unruh effect using uniformly accelerated two-level atomic probes. The acceleration-dependent factor generated during the atom--field interaction is encoded in the atomic state and can be estimated through repeated quantum measurements. Under a fixed total probe time, which represents the available measurement resource, we optimize the interrogation time, initial atomic state, and measurement basis of each probe to determine the optimal strategy for minimizing the estimation uncertainty.
Our results show that the achievable precision is governed by the Fisher information accumulated per unit probe time. By optimizing the evolution time of individual probes under a fixed total probing time, we demonstrate that shorter interrogation times allow for more sequential measurements and enhance the Fisher information accumulated per unit probing time, leading to improved estimation precision. We further reveal that the optimal initial state depend on both the acceleration factor and the probe evolution time. In particular, the excited state provides superior sensitivity in the weak-acceleration and small probe time regime.

Furthermore, we determine the minimum total probing time required to resolve the acceleration-dependent signal associated with the Unruh effect. The required measurement resource is strongly influenced by the atomic transition properties, including the transition frequency and spontaneous emission rate. By employing atomic systems with larger transition dipole moments, such as Rydberg atoms, the required total probing time can be substantially reduced.
These results demonstrate that quantum metrological optimization provides an effective strategy for probing acceleration-induced quantum effects and establishes a systematic framework for designing atomic experiments aimed at investigating the Unruh effect.

%When we estimate the parameters of initial atomic state, we find that the precision limits are controlled by the distance between atom and boundary. With proper atomic polarization direction and distance, the decay rates can become smaller than those in the unbounded vacuum case. The decay rates approach zero when the distance approaches zero with the polarization direction paralleling to the boundary.
%For the estimation of the atomic frequency, there exist a maximum quantum Fisher information and optimal measurement time, which can also be enhanced or decreased as compared to those in the case of flat space-time.

\begin{acknowledgments}
This work was supported by the National Natural Science Foundation of China under Grants No. 12165003, the special funding of talent program in Guizhou province[GCC[2023]005].
\end{acknowledgments}

\section*{Declarations}
\subsection*{Funding}
This work was supported by the National Natural Science Foundation of China under Grants No. 12165003, the special funding of talent program in Guizhou province[GCC[2023]005].
\section*{Availability of data and material}
My manuscript has no associated data. Data sharing is not applicable to this article as no new data were created or analyzed in this study.
\section*{Competing interests}
The authors declare that they have no competing interests.
\section*{Authors Contributions}
Not applicable. There is only one author.
%\appendix
%\section{the analytical result}
%Our interesting is asymptotic behavior of function

\end{document}